# Hot electron heatsinks for microwave attenuators below 100 mK

Jen-Hao Yeh,[1,2,a)] Rui Zhang,[1,2] Shavindra Premaratne,[1,2] Jay LeFebvre,[3,b)] F. C. Wellstood,[2,4] and B. S. Palmer[1,2]

[1]*Laboratory for Physical Sciences, 8050 Greenmead Drive, College Park, Maryland, 20740, USA*
[2]*Department of Physics, University of Maryland, College Park, Maryland, 20742, USA*
[3]*Department of Physics and Astronomy, University of California, Riverside, California, 92521, USA*
[4]*Joint Quantum Institute and Center for Nanophysics and Advanced Materials, University of Maryland, College Park, Maryland, 20742, USA*

We demonstrate improvements to the cooling power of broad bandwidth (10 GHz) microwave attenuators designed for operation at temperatures below 100 mK. By interleaving 9-µm thick conducting copper heatsinks in between 10-µm long, 70-nm thick resistive nichrome elements, the electrical heat generated in the nichrome elements is conducted more readily into the heatsinks, effectively decreasing the thermal resistance between the hot electrons and cold phonons. For a 20 dB attenuator mounted at 20 mK, a minimum noise temperature of $T_n \sim 50$ mK was obtained for small dissipated powers ($P_d < 1$ nW) in the attenuator. For higher dissipated powers we find $T_n \propto P_d^{1/4.4}$, with $P_d = 100$ nW corresponding to a noise temperature of 90 mK. This is in good agreement with thermal modeling of the system and represents nearly a factor of 20 improvement in cooling power, or a factor of 1.8 reduction in $T_n$ for the same dissipated power, when compared to a previous design without interleaved heatsinks.

a) Electronic mail: davidyeh@umd.edu
b) This research was performed while Jay LeFebvre was at Department of Physics, University of Maryland, College Park, Maryland, 20742, USA.

The field of superconducting quantum computing has advanced to a point where systems with multiple qubits are now available.[1,2] Essential advances that have enabled the development of multi-qubit systems include improvements in the energy relaxation time $T_1$ and coherence time $T_2$, which have risen by five orders of magnitude over the past two decades.[3,4] As $T_1$ increases, $T_2$ becomes increasingly sensitive to the presence of dephasing.[5–7] A common source of dephasing in superconducting qubits that use a circuit quantum electrodynamics (QED) architecture[8] is fluctuations in the number of photons in the cavity used to either measure the state of the qubit or act as a quantum bus.[6,9–11] Since remarkably small fluctuations in photon number can cause significant dephasing rates, it is essential to isolate the cavity from sources of thermally generated photons in the input and output control lines.[12]

To thermalize the microwave control lines connected to a superconducting circuit QED cavity, a series of microwave attenuators or directional couplers are commonly installed at the low-temperature stages in the measurement apparatus. These components reduce thermal noise from higher temperature stages, but when microwave signals are applied to control or read-out the device, electrical power will be dissipated in the resistive elements.[13] At millikelvin temperatures, a surprisingly small level of dissipated power can cause the temperature $T_e$ of the electrons in a thin-film normal metal element to increase well above the temperature $T_p$ of the phonons.[14,15] In the steady state, this "hot-electron" effect can be described by[15]

$$P_{e-p} = V\Sigma(T_e^5 - T_p^5) \qquad (1)$$

where $P_{e-p}$ is the rate at which heat flows from the electrons to the phonons, $V$ is the volume of the normal metal, and $\Sigma$ is the electron-phonon coupling constant, which is a material-dependent parameter. Since $P_{e-p}$ results in $T_e$ increasing above $T_p$, a resistive element that dissipates electrical power will produce more Johnson-Nyquist thermal noise, and cause more dephasing, than if it were not dissipating power. Hence, when designing an attenuator for operation at millikelvin temperatures, it is important to achieve a large effective coupling ($V\Sigma$) between the electrons and phonons, such that $T_e$ is close to $T_p$.[12] Since $\Sigma$ is fixed by the choice of material, this is most easily done by increasing the volume $V$ of the normal metal that is dissipating power.

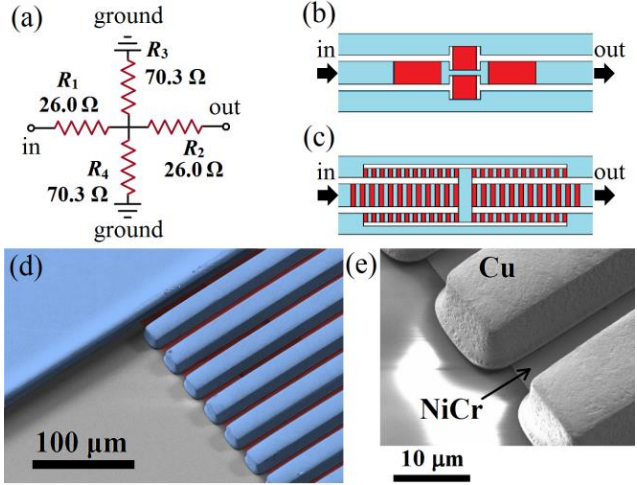

FIG. 1. (a) Circuit diagram for a single T-pad attenuator cell with 10 dB power attenuation and 50 Ω characteristic impedance. (b) Schematic drawing of a 10 dB cell for the non-interleaved design and (c) for the design with resistive elements interleaved by normal metal heatsinks. Red areas are resistive elements, and blue areas are conducting normal metal. Note: (b) and (c) are not drawn to scale, and (c) does not show the exact number of resistive elements or heatsinks. (d) False-colored scanning electron microscope (SEM) image of the Cu (blue) heatsinks on top of the NiCr (red) resistor and quartz substrate (gray). There is a 20-μm long Cu heatsink for each $L_R = 10$ μm long resistor. (e) SEM image of electroplated-Cu heatsinks with a thickness of 9 μm.

We recently reported results on heating in broadband microwave attenuators at cryogenic temperatures.[12] These attenuators were constructed from 10 dB attenuation cells with a T-pad configuration (see Fig. 1(a)) and had a characteristic impedance $Z_0 = 50$ Ω and a flat frequency response up to 10 GHz. The resistive elements in each cell were embedded in a coplanar waveguide geometry (Fig. 1(b)) and patterned from a sputtered $70 \pm 5$ nm thick nichrome (NiCr) film on a single-crystal quartz substrate, giving a sheet resistance $R_s = 25 \pm 5$ Ω/□. To connect the resistors together, a patterned highly conducting thin film of Ag was used. By using a transmon-cavity system as a sensitive average-photon-number thermometer,[16] a minimum noise temperature $T_n \lesssim 50$ mK was obtained at low dissipated powers $P_d$. However, for $P_d > 0.1$ nW, the output noise temperature of the attenuator increased as $T_n \propto P_d^{1/5.4}$. Both this power law and finite element simulations of the thermal response of the attenuator suggested that the heat was not efficiently conducted out of the nichrome, resulting in the electrons in the NiCr resistors being driven out of equilibrium with the phonons. If the thermal resistance between the electrons and the phonons could have been decreased, a lower noise temperature would have been achieved when dissipating power. In this article, we demonstrate how the cooling power of the attenuator can be improved by an order of magnitude by modifying the layout to increase the effectiveness of the hot-electron heatsinks.

First, to better conduct heat out of the NiCr resistor elements and into the cold conducting heatsinks,[15,17,18] we have divided the resistor elements in the attenuator circuit (see Fig. 1(b)) into many small sections (see Fig. 1(c)). The $L_R = 10$ μm length of the individual NiCr sections was chosen based on a detailed finite element simulation. We can understand what sets this length scale by examining the heat flow in a simplified geometry. In particular, consider two cold normal-metal heatsinks on each end of a resistor with length $L_R$ and cross-sectional area $A_R$. If all of the power $P_R$ dissipated in the resistor is conducted to and distributed uniformly in the heatsinks, the temperature of the electrons in the heatsinks will increase to

$$T_H \lesssim \left(T_p^5 + \frac{P_R}{V_H \Sigma_H}\right)^{\frac{1}{5}}, \qquad (2)$$

where $V_H$ is the total volume of the two heatsinks and $\Sigma_H$ is the electron-phonon coupling constant for the material used to construct the heatsinks. If we ignore the emission of phonons by the electrons in the nichrome (see supplementary material for a more complete discussion), then heat generated in the nichrome will only flow into the heatsinks by thermal conduction via the electrons, and the maximum electron temperature $T_R^{(max)}$ will be in the center of the resistor, furthest from the two heatsinks. For good cooling, $T_R^{(max)}$ should not be much larger than $T_H$, and we choose the specific temperature criterion $T_R^{(max)} < \sqrt{2} T_H$ for convenience. Using this criterion and Eq. (2), one finds an upper bound on the resistor length

$$L_R < \frac{4\kappa_e A_R}{P_R}\left(T_p^5 + \frac{P_R}{V_H \Sigma_H}\right)^{\frac{2}{5}}, \qquad (3)$$

where $\kappa_e T_e$ is the thermal conductivity of the resistor due to the electrons. For design values $T_p = 20$ mK, $P_R = 100$ nW, $\kappa_e = 0.02$ Wm⁻¹K⁻², $A_R = 56$ μm², $V_H = 1.4 \times 10^5$ μm³, and $\Sigma_H = 3 \times 10^7$ Wm⁻³K⁻⁵, we find $L_R < 10$ μm. The implication is that heat can be effectively removed from the NiCr electrons at low temperatures if the resistor sections have a length less than $L_R$. On the other hand, we require that the resistors in our T-pad attenuator circuit have specific values (see Fig. 1(a)). To achieve this, for $R_1$ and $R_2$ we used 83 sections of 10-μm long resistive NiCr elements interleaved with highly-conductive large volume heatsinks (see Fig. 1(d)), while $R_3$ and $R_4$ each had 208 interleaved resistor elements.

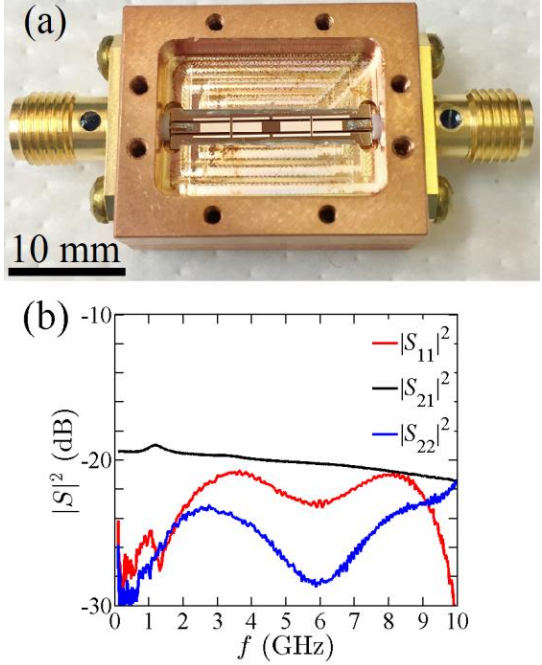

FIG. 2. (a) Photo of 20 dB attenuator packaged in a Cu box (lid not shown). (b) Magnitude of the scattering parameters $|S_{11}|^2$ (red), $|S_{21}|^2$ (black), and $|S_{22}|^2$ (blue) for the 20 dB attenuator.

The second change to the attenuator involved increasing the volume $V_H$ of the hot-electron heatsinks. Unfortunately, from Eq. (2) one can see that the electron temperature depends very weakly on the volume. To greatly increase the heatsink volume, thick Cu heatsinks were grown via electroplating.[19] Before performing the electroplating, the individual NiCr elements were covered with a patterned layer of photoresist. A 300 nm thick Cu seed layer was then electron-beam evaporated over the entire wafer, providing good electrical conduction to the electrodes. Next, an 11 μm thick photoresist layer was patterned to define the areas for Cu electroplating. After electroplating, the two photoresist layers were removed and the exposed regions of the Cu seed layer were etched away. Each resulting Cu heatsink had an average thickness of 9 μm and a lateral length of 20 μm between the NiCr elements (see Fig. 1(e)).

For testing, a 20 dB attenuator chip was packaged in a copper box (see Fig. 2(a)). The back of the chip and the two ground planes were secured to the box using silver epoxy. Room temperature measurements of the transmission $|S_{21}|^2$ of the packaged device showed an attenuation of $(20 \pm 2)$ dB from 0 to 10 GHz, with low levels of reflection $|S_{11}|^2$ and $|S_{22}|^2$ from the input and output ports (see Fig. 2(b)).

To measure the effective noise temperature of the attenuator at millikelvin temperatures, we connected the attenuator to the input line of an Al microwave cavity that contained a transmon qubit.[20] The transmon had a transition frequency $\omega_q/2\pi = 3.67$ GHz and consisted of a single Al/AlO$_x$/Al Josephson junction on a sapphire substrate. The three-dimensional superconducting aluminum cavity had a resonance frequency $\omega_c/2\pi = 7.96$ GHz. The transmon-cavity system and 20 dB attenuator were mounted on the mixing chamber stage of a dilution refrigerator with a 20 mK base temperature. To thermalize the signals from the 300 K stage to the tested attenuator, an additional 20 dB attenuator was mounted on the 3 K stage, and a 30 dB attenuator on the 70 mK stage (see Fig. 3(a)).[12] In a separate test for higher dissipated power (discussed below), the 30 dB attenuator was replaced with a 20 dB one. We extracted the qubit dephasing rate $\Gamma_\varphi$ and then used $\Gamma_\varphi$ to estimate the average photon number $\bar{n}_{th}$ stored in the cavity.[11,12,16] To increase the sensitivity of the qubit to thermal noise from the 20 dB attenuator, we made the coupling strength of the input cavity port larger than the output port (i.e. $Q_{in} = 1.8 \times 10^4 \ll Q_{out} = 7.9 \times 10^4$).

To find $\Gamma_\varphi$, we measured the qubit relaxation time $T_1$ and spin-echo coherence time $T_2$. The qubit dephasing rate $\Gamma_\varphi$ was then extracted using $\Gamma_\varphi = T_2^{-1} - (2T_1)^{-1}$. When no additional heating was applied to the attenuator, we measured mean values of $\langle T_1 \rangle = 14.2 \pm 0.1$ μs and $\langle T_2 \rangle = 26.7 \pm 0.8$ μs, where the error bars are the standard deviations of the mean of the 5 measurements. These results led to $\langle \Gamma_\varphi \rangle = (2 \pm 1) \times 10^3$ s$^{-1}$. By using the relation[11]

$$\Gamma_\varphi = \frac{\kappa}{2} \text{Re}\left[\sqrt{\left(1 + \frac{2i\chi}{\kappa}\right)^2 + \frac{8i\bar{n}_{th}\chi}{\kappa}} - 1\right] \quad (4)$$

along with measurements of the dispersive shift $\chi/2\pi = -0.34$ MHz and the cavity decay rate $\kappa/2\pi = 0.7$ MHz, we found that in this case the average number of photons in the read-out cavity was $\bar{n}_{th} = (1.0 \pm 0.7) \times 10^{-3}$. Given the total 70 dB attenuation in the input line, $\bar{n}_{th}$ has a lower bound of $\sim 10^{-4}$. Our result of $\bar{n}_{th}$ is about 5 times larger than that recently reported in a system in which the microwave signal to a one-port cavity was filtered by a bandpass cavity attenuator.[21] For an attenuator with output noise temperature $T_n$, we expect $\bar{n}_{th} = \left[\exp\left(\frac{\hbar\omega_c}{k_B T_n}\right) - 1\right]^{-1}$ and find $T_n \leq 55$ mK, similar to our previous design.[12] We note that fluctuations in our measured $T_1$ and $T_2$ produce significant uncertainty in $\Gamma_\varphi$ and the relatively larger fluctuations in $T_2$ suggest that other sources of dephasing may be present.

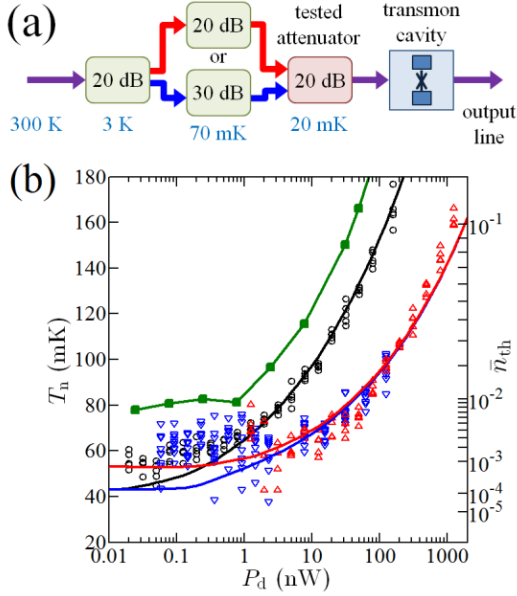

FIG. 3: (a) Schematic of attenuators along the input line and the typical temperatures of the stages where they are mounted. In separate cool-downs, a 20 dB or 30 dB attenuator was used at the 70 mK stage. (b) Effective noise temperature $T_n$ and average photon number $\bar{n}_{th}$ in the cavity versus dissipated power $P_d$ in a 20 dB attenuator. Red and blue triangles are data obtained when a 20 dB or 30 dB attenuator, respectively, was used at the 70 mK stage. Black circles are data from an attenuator with the previous design.[12] Red, blue, and black curves are from finite-element thermal simulations for the corresponding attenuator and input line layout. Green filled squares are data obtained on a commercial cryogenic attenuator (2082-6243-20-CRYO from XMA Corporation).

To evaluate the cooling power of the attenuator, we extracted the effective noise temperature $T_n$ while continuously dissipating power in the attenuator by applying a 1 GHz drive. Figure 3(b) shows $T_n$ versus the dissipated power $P_d$ for the 20 dB attenuator in two different cool-downs (blue and red triangles), our previous 20 dB cryogenic attenuator[12] (black circles), and a 20 dB commercial cryogenic attenuator (2082-6243-20-CRYO) from XMA Corporation (green squares). To measure the attenuator at higher dissipated powers (red triangles), we reduced the total attenuation in the refrigerator by replacing a 30 dB attenuator on the 70 mK cold plate with a 20 dB one. We also note that a traveling wave parametric amplifier[22] was used in the output line to amplify the cavity read-out signals while collecting the low power data set (blue triangles). Examining Fig. 3(b), one sees that it takes about 100 nW of dissipated power to drive the new attenuator to 100 mK. This power is about 10 times larger than the power required to drive the old version (black circles) to 100 mK, and 50 times larger than that of the commercial attenuator (green squares). The amount of improvement between our two attenuators is mainly due to the total volume $V_H$ of the new heatsinks being 260 larger than that of the previous design.

Finally, we can use the data in Fig. 3(b) to extract the electron-phonon coupling strengths of different materials. In our previous design without the interleaved heatsinks, simulations showed that the heat flow out of the NiCr resistors was the major bottleneck in cooling the electrons, so the output noise temperature was sensitive to the electron-phonon coupling strength of NiCr, and $\Sigma_{NiCr} = 5 \times 10^8$ Wm$^{-3}$K$^{-5}$ was found (see the black curve in Fig. 2(b)).[12] For the new design, where we anticipate more of the dissipated heat generated in the NiCr resistors to conduct into the heatsinks, we are more sensitive to the coupling of the electrons with the phonons in the Cu heatsinks. Setting the electron-phonon coupling strength of Cu to $\Sigma_{Cu} = 3 \times 10^7$ Wm$^{-3}$K$^{-5}$ yields good agreement between our finite-element simulations and the data (see blue and red curves in Fig. 3(b)). We note that this value of $\Sigma_{Cu}$ is comparable to previous reports, $\Sigma_{Cu} \approx 10^8$ Wm$^{-3}$K$^{-5}$.[15] This suggests that additional increases in the cooling power can be achieved by further enlarging the volume of the Cu heatsinks.

In summary, we have developed microwave attenuators with increased cooling power that are suitable for use with quantum devices below 100 mK. This improvement was achieved by decreasing the length that dissipated heat has to travel in each resistor element and interleaving large-volume hot-electron heatsinks. These simple components and related techniques for improving thermalization of signals at millikelvin temperatures will be useful for increasing the dephasing times of superconducting qubits, and possibly improving the performance of other types of millikelvin devices.[23–25]

## SUPPLEMENTARY MATERIAL

See supplementary material for the analysis of heat flow in a simplified geometry in order to estimate the length $L_R$ of the resistor sections.

## ACKNOWLEDGMENTS

The authors gratefully acknowledge Warren Berk for assisting with fabrication of the Cu heatsinks, and MIT Lincoln Laboratory (Greg Calusine and W. D. Oliver) for providing a traveling wave parametric amplifier. F.C.W. acknowledges support from the Joint Quantum Institute and the State of Maryland through the Center for Nanophysics and Advanced Materials.

## REFERENCES

[1] G. Wendin, Reports Prog. Phys. **80**, 106001 (2017).

[2] J. M. Gambetta, J. M. Chow, and M. Steffen, NPJ Quantum Inf. **3**, 2 (2017).

[3] W. D. Oliver and P. B. Welander, MRS Bull. **38**, 816 (2013).

[4] M. Reagor, W. Pfaff, C. Axline, R. W. Heeres, N. Ofek, K. Sliwa, E. Holland, C. Wang, J. Blumoff, K. Chou, M. J. Hatridge, L. Frunzio, M. H. Devoret, L. Jiang, and R. J. Schoelkopf, Phys. Rev. B **94**, 014506 (2016).

[5] C. Rigetti, J. M. Gambetta, S. Poletto, B. L. T. Plourde, J. M. Chow, A. D. Córcoles, J. A. Smolin, S. T. Merkel, J. R. Rozen, G. A. Keefe, M. B. Rothwell, M. B. Ketchen, and M. Steffen, Phys. Rev. B **86**, 100506(R) (2012).

[6] A. P. Sears, A. Petrenko, G. Catelani, L. Sun, H. Paik, G. Kirchmair, L. Frunzio, L. I. Glazman, S. M. Girvin, and R. J. Schoelkopf, Phys. Rev. B **86**, 180504(R) (2012).

[7] F. Yan, S. Gustavsson, A. Kamal, J. Birenbaum, A. P. Sears, D. Hover, T. J. Gudmundsen, D. Rosenberg, G. Samach, S. Weber, J. L. Yoder, T. P. Orlando, J. Clarke, A. J. Kerman, and W. D. Oliver, Nat. Commun. **7**, 12964 (2016).

[8] A. Blais, R.-S. Huang, A. Wallraff, S. M. Girvin, and R. J. Schoelkopf, Phys. Rev. A **69**, 062320 (2004).

[9] D. I. Schuster, A. Wallraff, A. Blais, L. Frunzio, R.-S. Huang, J. Majer, S. M. Girvin, and R. J. Schoelkopf, Phys. Rev. Lett. **94**, 123602 (2005).

[10] P. Bertet, I. Chiorescu, G. Burkard, K. Semba, C. J. P. M. Harmans, D. P. DiVincenzo, and J. E. Mooij, Phys. Rev. Lett. **95**, 257002 (2005).

[11] A. A. Clerk and D. W. Utami, Phys. Rev. A **75**, 042302 (2007).

[12] J.-H. Yeh, J. LeFebvre, S. Premaratne, F. C. Wellstood, and B. S. Palmer, J. Appl. Phys. **121**, 224501 (2017).

[13] B. Suri, Z. K. Keane, R. Ruskov, L. S. Bishop, C. Tahan, S. Novikov, J. E. Robinson, F. C. Wellstood, and B. S. Palmer, New J. Phys. **15**, 125007 (2013).

[14] M. L. Roukes, M. R. Freeman, R. S. Germain, R. C. Richardson, M. B. Ketchen, and S. State, Phys. Rev. Lett. **55**, 422 (1985).

[15] F. C. Wellstood, C. Urbina, and J. Clarke, Phys. Rev. B **49**, 5942 (1994).

[16] R. J. Schoelkopf, A. A. Clerk, S. M. Girvin, K. W. Lehnert, and M. H. Devoret, in *Quantum Noise Mesoscopic Phys.*, edited by Y. V. Nazarov (Springer Netherlands, 2003), pp. 175–203.

[17] R. C. Ramos, M. A. Gubrud, A. J. Berkley, J. R. Anderson, C. J. Lobb, and F. C. Wellstood, IEEE Trans. Appiled Supercond. **11**, 998 (2001).

[18] J. Pleikies, O. Usenko, R. Stolz, L. Fritzsch, G. Frossati, and J. Flokstra, Supercond. Sci. Technol. **22**, 114007 (2009).

[19] J. Gobet, F. Cardot, J. Bergqvist, and F. Rudolf, J. Micromechanics Microengineering **3**, 123 (1993).

[20] H. Paik, D. I. Schuster, L. S. Bishop, G. Kirchmair, G. Catelani, A. P. Sears, B. R. Johnson, M. J. Reagor, L. Frunzio, L. I. Glazman, S. M. Girvin, M. H. Devoret, and R. J. Schoelkopf, Phys. Rev. Lett. **107**, 240501 (2011).

[21] Z. Wang, S. Shankar, Z. K. Minev, P. Campagne-Ibarcq, A. Narla, and M. H. Devoret, *"Cavity Attenuators for Superconducting Qubits,"* submitted to Phys. Rev. Applied.

[22] C. Macklin, K. O'Brien, D. Hover, M. E. Schwartz, V. Bolkhovsky, X. Zhang, W. D. Oliver, and I. Siddiqi, Science **350**, 307 (2015).

[23] P. K. Day, H. G. LeDuc, B. A. Mazin, A. Vayonakis, and J. Zmuidzinas, Nature **425**, 817 (2003).

[24] C. Enss, *Cryogenic Particle Detection* (Springer, 2005).

[25] F. Giazotto, T. T. Heikkilä, A. Luukanen, A. M. Savin, and J. P. Pekola, Rev. Mod. Phys. **78**, 217 (2006).